# Novel Remapping Approach for HR-EBSD based on Demons Registration

Chaoyi Zhu[a], Kevin Kaufmann[b], Kenneth S. Vecchio[a,b]

[a]Materials Science and Engineering Program, UC San Diego, La Jolla, CA 92093, USA

[b]Department of NanoEngineering, UC San Diego, La Jolla, CA 92093, USA

**Abstract**

In this study, the possibility of utilizing a computer vision algorithm, i.e., demons registration, to accurately remap electron backscatter diffraction patterns for high resolution electron backscatter diffraction (HR-EBSD) applications is presented. First, the angular resolution of demons registration is demonstrated to be lower than the conventional cross-correlation based method, particularly at misorientation angles > 9°. In addition, GPU acceleration has been applied to significantly boost the speed of iterative registration between a pair of patterns with 10° misorientation to under 1s. Second, demons registration is implemented as a first-pass remapping, followed by a second pass cross-correlation method, which results in angular resolution of $\sim 0.5 \times 10^{-4}$ degrees, a phantom stress value of ~35MPa and phantom strain of $\sim 2 \times 10^{-4}$, on dynamically simulated patterns, without the need of implementing robust fitting or iterative remapping. Lastly, the new remapping method is applied to a large experimental dataset collected from an as-built additively-manufactured Inconel 625 cube, which shows significant residual stresses built-up near the large columnar grain region and regularly arranged GND structures.

1. Introduction

Electron backscatter diffraction (EBSD) has attracted many users over the years as it becomes part of a standard laboratory facility. EBSD allows users to quickly gain rich information from crystalline materials such as crystallographic orientation, microstructure, defect density, and (relative) residual stress/strain [1]. In recent years, high-resolution EBSD (HR-EBSD) has been an active research area in the EBSD community because of its potential to accurately map residual stress/strain with high sensitivity. Since EBSD is an orientation measurement technique, improving the angular resolution of the technique is of most interest to researchers. Methods to improve angular resolution have been developed such as the pattern comparison method [2], iterative indexing [3], the 3D-Hough transform [4] and HR-EBSD [5]. In particular, the high resolution (or high angular resolution) electron backscatter diffraction (EBSD) technique is capable of characterizing additional information such as residual stress/strain [6]. Moreover, it is conveniently based on a scanning electron microscope, part of a standard laboratory facility, equipped with an electron backscatter detector. Compared to neutron diffraction or high energy X-ray diffraction approaches to measure residual strain/stress, HR-EBSD would be a much more attractive method if some of the challenges in HR-EBSD can be fully resolved.

There are basically two main challenges in HR-EBSD. First, the absolute strain or stress state of the reference pattern in the HR-EBSD method is an unknown state. Therefore, the residual stress or strain in the current HR-EBSD framework is a relative quantity. Simulation based approaches to determine the strain state of the reference patterns is certainly a promising solution [7,8], but the uncertainties in the experimental setup and pattern center accuracy will limit the sensitivity [6]. A recent study by Tanaka *et al.* [9] has utilized a differential evolution method to optimize measured Euler angles and pattern center of the experimental reference pattern, which demonstrates promising sensitivity for dynamical simulation based 'absolute' HR-EBSD. In addition, full-field integrated DIC method developed by Vermeij *et al*. [10] also tries to answer the question of solving 'absolute' residual deformation without the need of a zero-strain reference pattern using only dynamically simulated patterns.

Second, metals under deformation usually have large lattice rotation variation within a grain. Determination of the position of the cross-correlation function (XCF) peak is therefore prone to error because significant lattice rotation tends to skew the shape of the XCF peak or even produce multiple peaks, which requires a remapping technique to account for the phantom strain [11,12]. The two current remapping techniques are based on 1) a disorientation matrix obtained from orientation matrices, or 2) first-pass rotation estimate with cross-correlation, both of which have some areas for possible improvement.

The remapping method based on the disorientation matrix, obtained from the Hough transform, has the advantage of being very fast and more accurate at significantly large misorientations (>10°), but it has limited accuracy at small angles by the uncertainty in the rotation axis [12]. In the StrainCorrelator developed by Maurice *et al*. [12], cross-correlation with and without remapping are both carried out, and the one with a lower mean angular error is selected. In addition, this type of remapping is also limited by the angular resolution of the Hough-based EBSD, typically around 0.5° to 1° [13–16], given a precise geometry of the scan setup. Symmetry operators must also be considered to deal with 'sudden jumps' in the indexing software due to crystallographically equivalent Euler angle sets.

The other remapping technique is based on a first-pass cross-correlation to estimate the rotation part, and then apply a second pass to extract the residual deformation between the remapped pattern and test pattern. Its angular resolution at lower misorientation angles is certainly much improved compared to the previous method, but the first-pass cross-correlation, without iterative robust fitting [17], does not extract an accurate rotation matrix when lattice rotation is significant, which might require a second remapping. Hence, the need for higher accuracy comes at the expense of more computational power, which limits the use of HR-EBSD to very small data sets.

Recently, new global image cross-correlation based techniques have been proposed to simultaneously improve the strain and rotation sensitivity of HR-EBSD [10,18,19], inspired by the studies on matching the whole region of interest [20–24]. In this work, we propose a new remapping method based on an image registration method (i.e. demons registration) [25]. The first demons registration paper was published by Thirion *et al.* [25] as an analogy of Maxwell's demon and has later been widely implemented for many radiotherapy applications [17–21]. The demons registration approach resembles the global image cross-correlation technique in that it also maps the non-rigid transformation of the images. In the demons registration algorithm, the displacement field to register the undeformed (f) pattern to deformed (m) pattern is computed through the optical flow equation (given by Eq. 1):

$$\vec{u} = \frac{[I_m(\vec{r}_m) - I_f(\vec{r}_f)]\nabla I_f(\vec{r}_f)}{\nabla I_f(\vec{r}_f)^2 + [I_m(\vec{r}_m) - I_f(\vec{r}_f)]^2} \quad (1)$$

where $\vec{u}$ is the displacement field, $I_m(\vec{r}_m)$ is the intensity of a point in the deformed pattern at a position $\vec{r}_m$, $I_f(\vec{r}_f)$ is the intensity of point in the undeformed pattern at a position $\vec{r}_f$, and $\nabla I_f(\vec{r}_f)$ is the intensity gradient of the deformed image at a position $\vec{r}_f$. The displacement field is calculated iteratively with the demons registration.

For large deformation, the multi-resolution demons algorithm is more robust in terms of mapping the accurate displacement field [26]. The multi-resolution approach essentially uses a displacement field obtained from the downscaled version (higher pyramid level) of the image as input displacement for the next higher resolution version (lower pyramid level) of the same image to better capture the large-scale feature variation. In this study it is shown that by applying the multi-resolution demons registration alone does not have enough sensitivity for accomplishing HR-EBSD. It is, however, possible to use multi-resolution demons registration to first remap the reference pattern, with similar accuracy compared to cross-correlation based remapping over a larger span of rotation angles, before using conventional cross-correlation to obtain rotation and strain sensitivity down to $0.5 \times 10^{-4}$ and $2 \times 10^{-4}$; respectively. It is important to note that the accuracy of strain is influenced by many other factors in addition to lattice rotation [6], which should all be appropriately addressed at the same time to obtain a physically plausible strain level when this approach is applied to experimental patterns.

## 2. Methodology

The conventional HR-EBSD technique is based on the use of cross-correlation of many regions of interests (ROIs) [5,27] and a convolution process between two functions in the Fourier domain, to extract shifts, $\vec{q}$, between two patterns. This is an inverse Fourier transform of the element-wise product of the test pattern (deformed) in its frequency space and the conjugate of the reference pattern (assumed to be undeformed) in its frequency space. In practice, the intensity of the pattern image is first normalized to have a mean of zero. The Hanning window [5] is then applied to each of the ROIs before transforming into Fourier space to eliminate the image edge effect. Then the peak position relative to the origin of the XCF can be determined with subpixel sensitivity by interpolating the XCF peak.

The shifts at the center of each ROI can alternatively be obtained through the multi-resolution demons registration, which iteratively determines the displacement field required to transform a reference pattern toward the test pattern. The displacement field near the edges of the pattern needed to be carefully avoided since these regions contain erroneous displacement fields due to lack of similarity.

The relationship between the shift of ROIs (center) and deformation gradient tensor or displacement gradient tensor can then be mathematically formulated. A formulation used by Maurice *et al*. [12] is adopted here to calculate the deformation gradient tensor associated with the shifts in patterns.

$$\vec{r'} = \vec{r} + \vec{q} = \frac{Z^*}{(\boldsymbol{F} \cdot \vec{r}) \cdot \vec{k}} \boldsymbol{F} \cdot \vec{r} \qquad (2)$$

where $\vec{r}$ is the position vector of the center of region of interests in the reference pattern, $\vec{q}$ is the measured shift vector in the detector frame, $\vec{r'}$ is the position vector of the deformed center of region of interests, $Z^*$ is the detector distance from the source point to the pattern centre, $\vec{k}$ is a unit vector normal to the screen, and $\boldsymbol{F}$ is the deformation gradient tensor to be solved. A schematic diagram is shown below in Fig. 1 to visualize the quantities mentioned above and the associated coordinates systems.

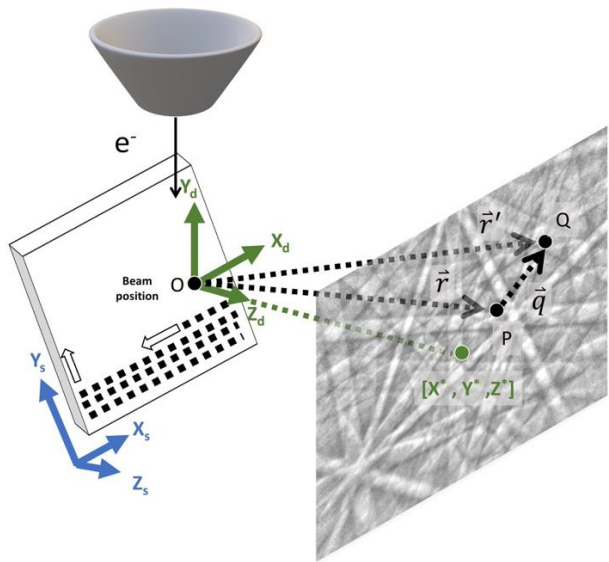

Fig. 1: Schematic of EBSD setup and the quantities used in HR-EBSD.

A set of circular regions of interests are placed evenly spaced around the pattern center plus one at the pattern center to measure the necessary shift vectors $\vec{q}$ through cross-correlation to establish an overdetermined system of equations. In this paper, the center positions of ROIs employed in the cross-correlation are used to extract shifts obtained from the demons registration. The deformation gradient tensor can then be solved using non-linear least square minimization algorithm under the equality constraint defined by the traction free boundary condition.

$$minf(F) = \sum_{\{ROI\}} \frac{1}{2}\left|\frac{Z^*}{(F \cdot \vec{r}) \cdot \vec{k}} F \cdot \vec{r} - (\vec{r} + \vec{q})\right|^2 \tag{3}$$

To extract the rotation part of the deformation gradient tensor, a singular value decomposition method is used ($P$ and $Q$ are rotation matrices, $D$ is a stretch matrix).

$$F = PDQ^T \tag{4}$$

Therefore, the total rotation matrix can be expressed in terms of $P$ and $Q$.

$$R = PQ^T \tag{5}$$

It is then possible to express the rotation matrix in axis-angle pair notation. The rotation axis $m = [m_1, m_2, m_3]$ of the rotation matrix $R$ is the eigenvector when the corresponding eigenvalue of R is equal to unity; the rotation angle $\theta$ around the rotation axis is given by $\theta = acos[(tr(R) - 1)/2]$. The lattice rotation tensor is related to the axis-angle pair relation through:

$$\omega_{ij} = -\varepsilon_{ijk} m_k \theta = -\varepsilon_{ijk} \theta_k \tag{6}$$

where $\theta_k$ is the rotated vector given by $\theta_k = m_k \theta$.

Writing out the lattice rotation tensor in full gives the antisymmetric matrix in the following:

$$\boldsymbol{\omega} = \begin{pmatrix} 0 & -\theta_3 & \theta_2 \\ \theta_3 & 0 & -\theta_1 \\ -\theta_2 & \theta_1 & 0 \end{pmatrix} = \begin{pmatrix} 0 & -\omega_{12} & \omega_{31} \\ \omega_{12} & 0 & -\omega_{23} \\ -\omega_{31} & \omega_{23} & 0 \end{pmatrix} \tag{7}$$

From the elastic rotation tensor $\Omega_k^e$ (antisymmetric part of the elastic distortion tensor), the lattice rotation tensor is related to the orientation matrix through Eq. 8.

$$\Omega_k^e = -\frac{1}{2}e_{ijk}\omega_{ij} = -\frac{1}{2}e_{ijk}(g_{ij}-\delta_{ij}) \tag{8}$$

Therefore, the components of the lattice rotation tensor can also alternatively be obtained from the orientation matrix **g**:

$$\omega_{23} \approx \frac{1}{2}(g_{23}-g_{32}) \tag{9}$$

$$\omega_{31} \approx \frac{1}{2}(g_{31}-g_{13}) \tag{10}$$

$$\omega_{12} \approx \frac{1}{2}(g_{12}-g_{21}) \tag{11}$$

With or without remapping, the strain tensor and Cauchy stress tensor can be obtained using Eq. 11 and Eq. 12. Since the elastic strain (<0.2%) is rather small, the Green-Lagrange strain is approximately equal to the infinitesimal strain. If the remapping technique is applied using a cubic interpolation scheme, the deformation gradient tensor **F** will be further decomposed into a finite rotation matrix $\boldsymbol{R}_{finite}$ (pre-determined by Equations 3-5) and an infinitesimal deformation gradient tensor $\boldsymbol{F}_{infinitesimal}$ calculated between the remapped reference pattern (using $\boldsymbol{R}_{finite}$) and the test pattern [12].

$$\boldsymbol{F} = \boldsymbol{F}_{infinitesimal}\boldsymbol{R}_{finite} \tag{12}$$

It is important to realize that the total deformation gradient tensor in this case is in the detector frame. Coordinate transformation can be readily established based on the tilt of the phosphor screen with respect to the sample coordinate system $F_S$.

$$\boldsymbol{R}_{\theta_{tilt}} = \begin{pmatrix} \cos\theta_{tilt} & \sin\theta_{tilt} & 0 \\ -\sin\theta_{tilt} & \cos\theta_{tilt} & 0 \\ 0 & 0 & 1 \end{pmatrix} \tag{13}$$

$$\boldsymbol{F}_S = \boldsymbol{R}_{\theta_{tilt}}\boldsymbol{F}\boldsymbol{R}_{\theta_{tilt}}^T \tag{14}$$

The total deformation gradient tensor in the sample frame can be used to calculate the strain tensor with Eq. 11. Using Hooke's law, the surface stress state (in the deformed state) can be obtained under the traction-free boundary condition [28] and crystal orientation (note that the stiffness matrix C needs to be rotated into the correct orientation using the orientation matrix [29], which is described in Supporting Materials Part 2). The elastic constants of nickel (in the Cartesian crystal frame) are taken from the Materials Project Website (https://materialsproject.org/): $C_{11}=2.76\times10^{11}$ Pa, $C_{12}=1.59\times10^{11}$ Pa, $C_{44}=1.32\times10^{11}$ Pa.

$$\varepsilon_S = \frac{1}{2}(F_S F_S^T - I) \approx \frac{1}{2}(F_S + F_S^T) - I \tag{15}$$

$$\sigma_S = C : \varepsilon_S \; (\vec{t} = \sigma_S Z_s = [0,0,0]) \tag{16}$$

## 3. Results and Discussion

### 3.1. HR-EBSD Part I: Angular Resolution

All programming in this paper is implemented in Matlab except the dynamically simulated patterns, which are simulated from EMsoft [30] for FCC nickel generated at 20 kV. The pattern resolution is [1244, 1024] pixels with the pattern center located at [0.5, 0.5, 0.5]. The sample tilt is 70° and detector tilt is 10°; the Euler angles used here adopt the HKL convention.

The first test set involves determining the accuracy of one component of the lattice rotation matrix $\omega_{12}$, the in-plane rotation. The reference pattern orientation, expressed in Euler angles, is [0°, 0°, 0°] ([0, 0, 0] in rad). The test patterns are simulated at orientations from [1°, 0°, 0°] to [10°, 0°, 0°] at 1° intervals ([0.0175, 0, 0] to [0.1745, 0, 0] in rad). For cross-correlation purpose, 100 ROIs are arranged in a circular pattern around the pattern center (plus one right at the pattern center) with ROI size of 256 pixels by 256 pixels, as shown in Fig. S1. The center of ROIs in Fig. S1 are used to obtain the shifts vectors from the displacement field (shift field) as shown in Fig. S2.

Moreover, two methods are used to extract finite rotation for the remapping technique: 1) the cross-correlation method, and 2) the demons registration method, as shown in Fig. S3. In addition, the image registration parameters used here are iterations = 25 (at each pyramid level), pyramid level = 7, and no filtering or smoothing is applied on the displacement field. The demons registration converges to the correct rotation tensor at around 20 iterations (for each pyramid level × 7 pyramid levels = 140 iterations) for a computational time around 4s on an Intel i7-8750H CPU processor, see Fig. S4. GPU acceleration can also be implemented to greatly speed up the computational cost to lower than 1s, in this case, on a NIVIDIA GeForce GTX 1050 Ti GPU.

Figure 2 shows a comparison of the absolute rotation accuracy of the four methods: Fig. 2(a) cross-correlation, Fig. 2(b) demons registration, Fig. 2(c) $1^{st}$ pass remapping with cross-correlation + $2^{nd}$ pass cross-correlation, and Fig. 2(d) $1^{st}$ pass remapping with demons registration + $2^{nd}$ pass cross-correlation. It is clear that the demons registration method in Fig. 2(b) measures the rotation components more accurately for the applied rotation $\omega_{12}$ compared to the cross-correlation method in Fig. 2(a), especially at large rotations. However, the demons registration does lead to a slightly higher rotation error in the $\omega_{23}$, which is likely associated with pattern distortion introduced by the sample tilt around the x axis. Overall, if these two methods were to be used for remapping, it would be expected that the demons registration would yield better performance. This is demonstrated in Fig. 2(d), in which the demons registration improves the rotation accuracy of the cross-correlation based remapping from $1.8 \times 10^{-4}$ to $0.5 \times 10^{-4}$.

The mean angular error (MAE) and peak height or structure similarity measure (PH or SSIM) for the four different methods are also compared in Fig. 3. The MAEs for cross-correlation and demons registration are similar at lower rotations (<9°), but the demons registration has much lower MAE at rotation >9°. Additionally, the remapping method based on demons registration is slightly better than the cross-correlation based remapping, which leads to lower rotation error in Fig. 2(d). In this study, the SSIM is only used for demons registration method because no cross-correlation is involved. The PH or SSIM values drop drastically for the methods without remapping whereas the two remapping methods both seem to retain very high values.

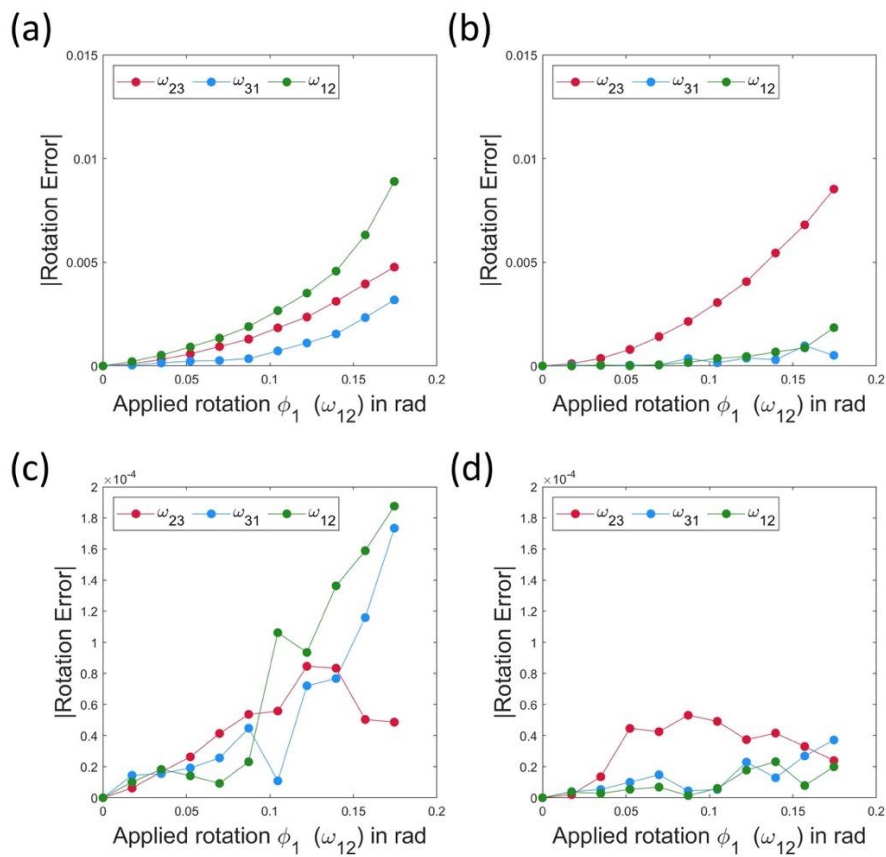

Fig. 2: Absolute rotation error of the measured lattice rotation tensor components with (a) cross-correlation, (b) demons registration, (c) 1st pass remapping with cross-correlation+2nd pass cross-correlation and (d) 1st pass remapping with demons registration+2nd pass cross-correlation. Note the different y-axes in a,b and c,d.

Second test set includes two simulated patterns of different orientations with three different Euler angles. The reference pattern orientation, expressed in Euler angles, is [125˚, 43˚, 57˚] ([2.1817, 0.7505, 0.9948] in rad). The test pattern orientation, expressed in Euler angles, is [130˚, 40˚, 60˚] ([2.2689, 0.6981, 1.0472] in rad). The misorientation angle between the two patterns is ~8˚. The convergence of demons registration to the expected rotation matrix requires approximately 10 iterations (for each pyramid level×7 pyramid levels=70 iterations) for a computational time around 2s on an Intel i7-8750H CPU processor and less than 1s on the NIVIDIA GeForce GTX 1050 Ti GPU. The image intensity difference between reference and test patterns are illustrated in Fig. 4(a).

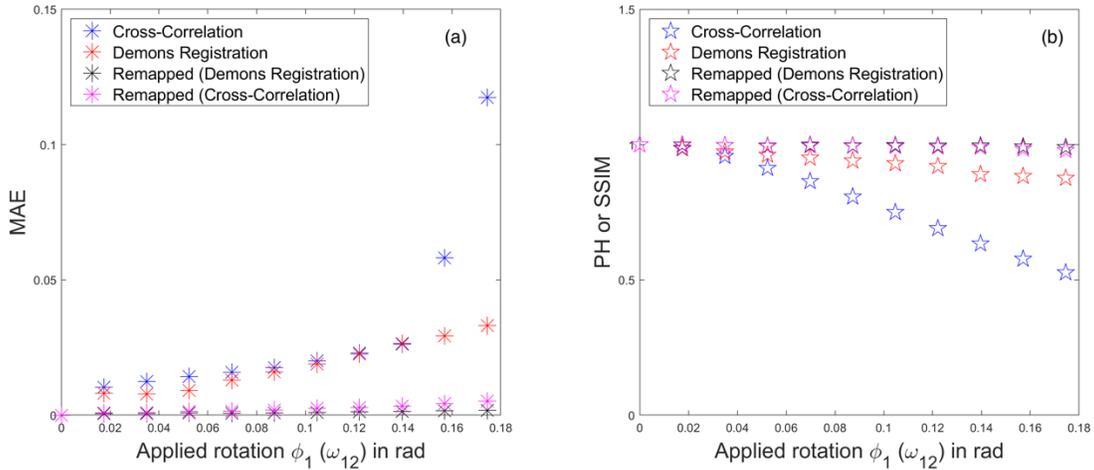

Fig. 3: (a) Mean angular error (MAE) and (b) peak height (PH) or structure similarity measure (SSIM) of four different HR-EBSD methods.

The same four methods of determining the rotation matrix are used as those in the first test set. With demons registration, the reference pattern can be warped to the test pattern as illustrated by the intensity difference plot in Fig. 4(b), in which the only unregistered part is the edge of the pattern. Based on the green part of the figure (edge of test pattern) in Fig. 4(b), the registered reference is clearly well aligned with the test pattern underneath it. As shown in Fig. 4(e), the absolute rotation error of the three components of lattice rotation tensor are approximately similar for the two methods without remapping ($\sim 10^{-3}$), whereas the two remapping methods can reduce the error by an order of magnitude. In addition, the remapping method based on demons registration slightly improves the cross-correlation based remapping approach.

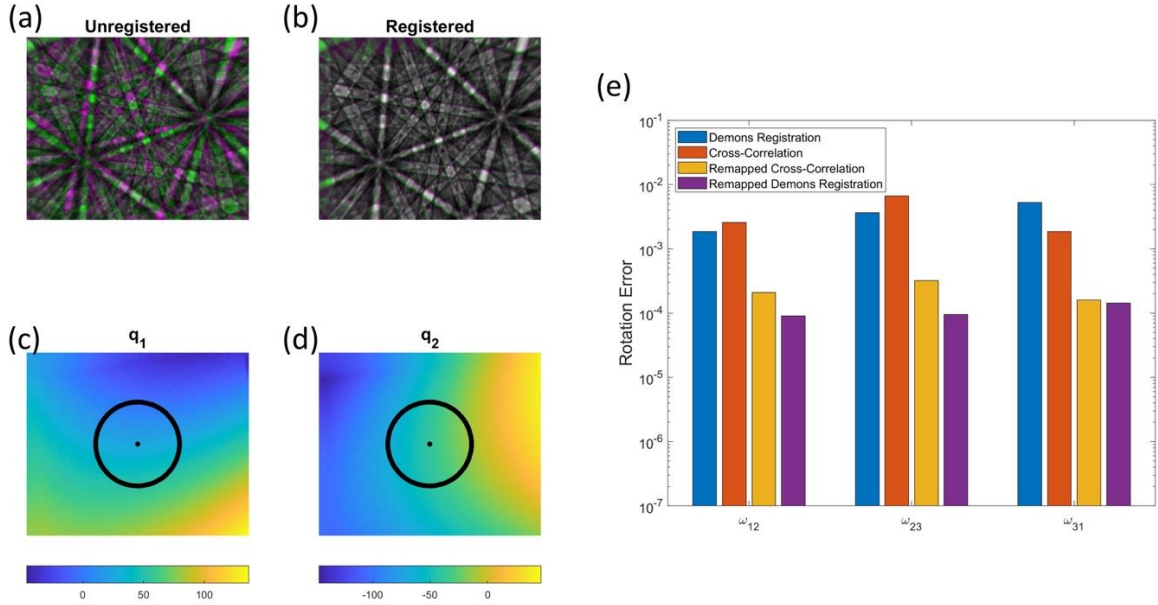

Fig. 4: (a) The unregistered and (b) registered intensity of patterns are plotted on top of the components of the calculated displacement fields (c) $q_1$ and (d) $q_2$, required to match the reference pattern towards test pattern. Black ring of dots is the 100 positions used to extract the shift vector $\vec{q} = [q_1, q_2]$, which are on the same positions as the center of ROIs for cross-correlation. The right bar figure (e) shows a comparison of the absolute rotation error (in rad) associated with the three components of lattice rotation tensor using four difference methods.

3.2 HR-EBSD Part II: Stress and Strain Sensitivity

The first test set is used again in validating strain sensitivity; strain and stress are calculated using Eq. 11 and Eq. 12. With only the applied rotation, the first test set should be stress and strain free; therefore, any non-zero values for the stress or strain is regarded as a phantom stress or strain. The phantom strain associated with large rotation is clearly shown in Fig. 5(a-d) for the four methods described previously. It is also clear that using demons registration alone results in significant phantom strain at the highest applied rotation angle ($\sim 2\times 10^{-3}$), which is lower than the compared to the cross-correlation based method ($\sim 5 \times 10^{-3}$), see Fig. 5(a-b). Moreover, the corresponding phantom stress can reach as high as 600MPa and 400MPa, nearly reaching the yielding point of some materials. Nevertheless, the typical elastic limit is around $2\times10^{-3}$, which means that neither cross-correlation nor demons registration can be employed alone for patterns with significant rotation. With the implementation of the remapping method, phantom strains have been effectively reduced to $\sim6\times10^{-4}$ and $\sim2\times10^{-4}$ for the two different remapping methods, as shown in Fig. 5(c-d). The improved strain error for demons registration based remapping is attributed to the more accurate remapping rotation matrices extracted especially at higher applied rotation angles. Additionally, the phantom strains for the two remapping methods at small applied rotations (<5°) are similar in magnitude $\sim1.5\times10^{-4}$. In Fig. 5(g-h), the phantom stresses for the two remapping methods are around 20MPa at small applied rotations (<5°). The phantom stress for the demons registration remapping method can remain as low as ~35MPa at the highest applied rotation whereas the cross-correlation based remapping method produces phantom stress of ~70MPa.

3.2. Shift and Zoom/Shrink Correction

To implement the HR-EBSD for analyzing real patterns, shift and zoom/shrink of an EBSP need to be corrected first. The pattern shift correction can be determined geometrically from the shift of beam position in the microscope reference frame and then transformed into the detector frame. In addition, the zoom and shrink factor can be determined using the ratio between the detector distance of the test and reference patterns and applied on either the test or reference pattern through an affine transformation.

A large test scan (200μm by 200μm) is obtained from an unstrained single crystal Si sample, collected by the Thermo Fisher (formally FEI) Apreo SEM equipped with an Oxford *Symmetry* CMOS EBSD Detector. The pattern center optimization step is carried out with the 'local optimize function' in Oxford Aztec software by capturing several patterns at different detector distances.

By using one experimental pattern at the top left corner as the reference pattern, every other pattern is cross-correlated with the reference pattern without the need for a remapping method. The distribution of phantom strains for shift and zoom corrected and uncorrected methods are shown in Fig. 6. The shift and zoom/shrink corrected patterns lead to significant reduction in the phantom strain of $\varepsilon_{11}$ and $\varepsilon_{22}$, which restricts the distribution of all the components of the strain tensor to fall within $\pm 0.001$. Most of the phantom strains (~70% area fraction) sit close to the zero position, but some off-zero phantom strains exist likely due to small uncertainties in the geometry of setup, e.g., sample/detector tilt angle, angle between sample surface normal and detector screen normal.

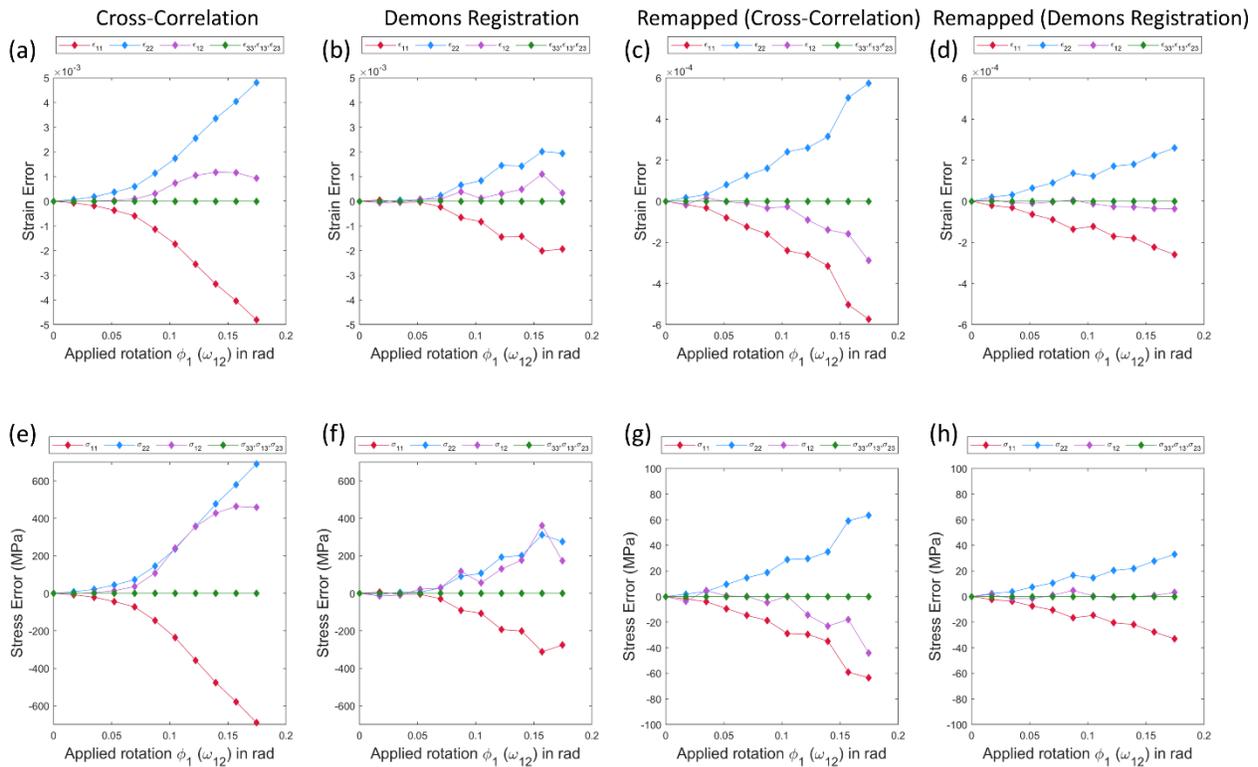

Fig. 5: Phantom strain (a-d) and stress (e-h) obtained using four different HR-EBSD methods.

3.3 Lattice rotation and residual stress/strain in additively manufactured Inconel 625

Additive manufacturing of metal components is a promising way of revolutionizing manufacturing because of its ability for fast prototyping, design of complex geometry, etc. [31–33]. However, the inherent problems of additive manufacturing, very similar to problems experienced in welding, are generation of defects and build-up of thermal stresses, which adversely affect the fatigue performance and fracture resistance [34,35]. Therefore, understanding of the types of defects and residual stress distribution in additively manufactured samples will assist in optimizing printing parameters, design strategies, and post-processing.

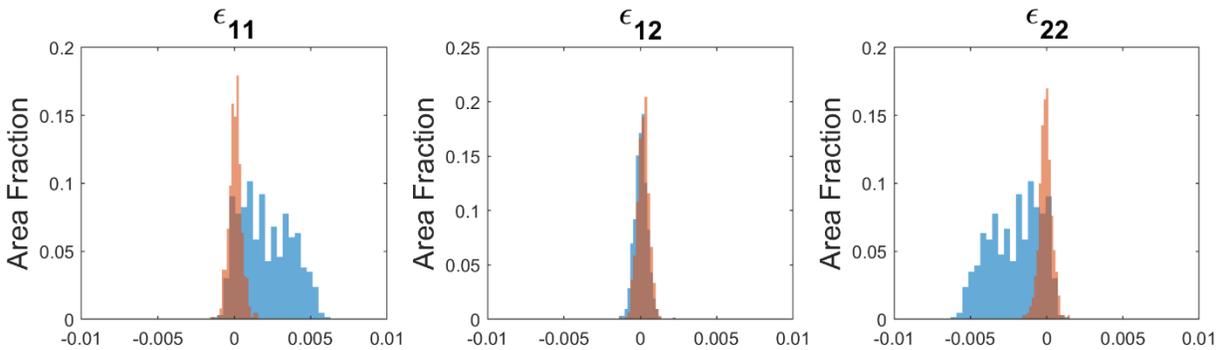

Fig. 6: Phantom strain on a single crystal Si sample obtained from a large area (200µm by 200µm)

With the use of HR-EBSD, defects such as geometrically necessary dislocations and residual stress can be mapped. In this part of the study, an additively manufactured Inconel 625 cube was printed in-house using a Formalloy L221 system and a NUBURU AO-150 blue laser. The cube was built in an argon atmosphere using the following deposition parameters: 120W laser power, 120µm slice height, 250µm hatch spacing, 1,000 mm/min infill speed, 800 mm/min contour speed, and a powder flow rate of ~1 g/min.

Selection of experimental reference patterns (red dots in Fig. 7) is automatically determined from the collected diffraction pattern data set for each grain, as shown in Fig. 7(a), after the grain structure has been segmented with the high angle grain boundary ~15˚. The selection metric is established based on the maximum ratio between normalized band contrast (Fig. 7(d)) and normalized mean angular deviation (MAD) within each grain. An example of a reference pattern obtained from the large columnar grain is shown in Fig. 7(b) with dynamic and static background correction applied to process all the patterns to enhance contrast of bands. In addition, Fig. 7(c) shows the Hough based misorientation values between test patterns and the selected reference pattern of each grain. Substructure in the columnar grain region is more pronounced compared to other grains, which represents an area of interest for further exploration.

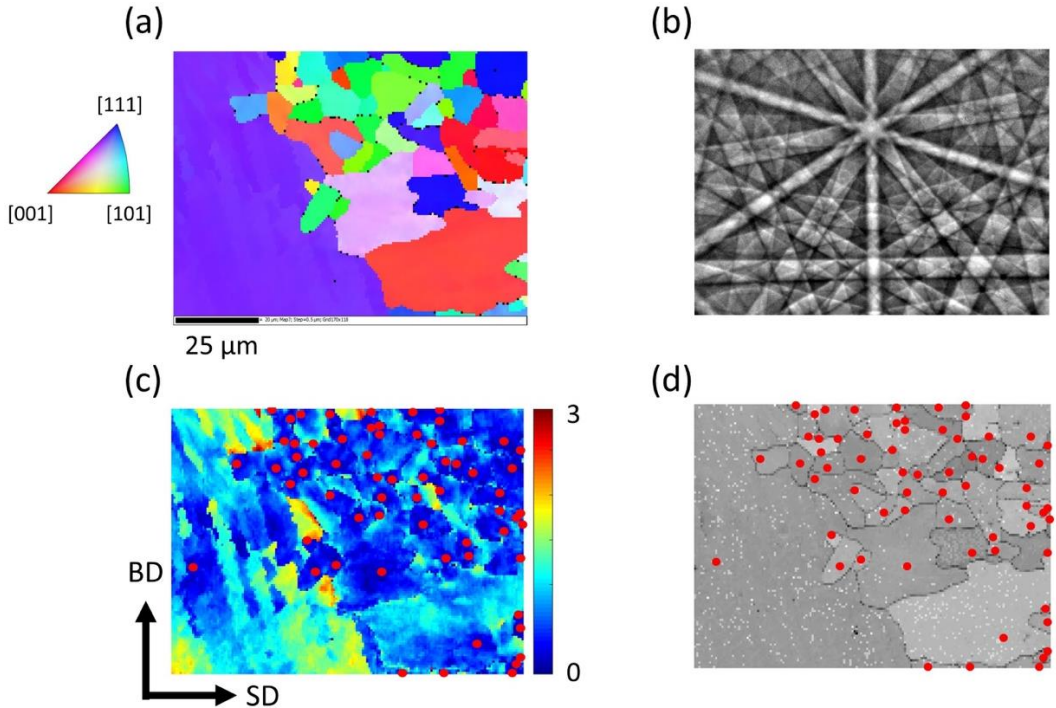

Fig. 7: (a) Inverse pole figure (BD: build direction, SD: scan direction); (b) background corrected (dynamic + static background correction) electron backscatter diffraction pattern; (c) reference misorientation map (deg); (d) band contrast map.

For the demons registration-based HR-EBSD analysis, a low pass filter and high pass filter are applied first to remove high frequency noise and low frequency background gradients. 45 ROI centers (256 by 256 pixels placed circularly around the pattern center at a radial distance of 250 pixels) are used to extract the $R_{finite}$ rotation matrix to remap the reference pattern toward the test pattern (number of pyramid levels = 7, number of iterations = 20/pyramid level). Cross-correlation between the remapped reference pattern and test pattern is then used to obtain the infinitesimal deformation gradient tensor. Combined with the finite rotation used for remapping, the total deformation tensor is calculated with Eq. (12). The resultant lattice rotation tensor from HR-EBSD can be compared with the Hough based lattice rotation tensor. Equations 4-7 is the singular value decomposition approach to obtain the rotation matrix and the associated lattice rotation tensor. The Hough based lattice rotation tensor is estimated from components of the orientation matrix according to Equations 8-11. As shown in Fig. 8, the lattice orientation tensor components are nominally similar. The line profiles indicate that $\omega_{23}$ matches very well for the two different methods whereas the HR-EBSD method shows smoother line profiles for $\omega_{12}$ and $\omega_{31}$ compared to the Hough-EBSD method. It is also interesting to note that the columnar grain shows slightly different absolute values of rotation components likely due to systematic error in obtaining the absolute orientation of the EBSD pattern from the Hough transform but the spatial variations match well.

From the total deformation gradient tensor, the local residual (relative) stress and strain can be calculated using Equations 15-16. Fig. 9 is a combined plot for lattice rotation, residual strain and residual stress obtained with the HR-EBSD analysis. The color scale of residual strain is defined by the typical elastic limit of a metal, ~0.002, and the color scale of residual stress is defined by the yield strength $\sigma_y$ of Inconel 625, ~500MPa. The distribution of residual stress or strain indicates that the columnar region has a significant build-up of thermal stresses ~0.5-0.8$\sigma_y$, which is likely due to the thermal gradient during the printing process. This level of residual stress is sufficient to lead to detrimental effects on the mechanical behavior of additively manufactured Inconel 625, which requires post-processing treatment such as annealing. In addition, stresses are also concentrated near grain boundaries and triple junctions for the necessary strain compatibility for the relatively equiaxed grain region.

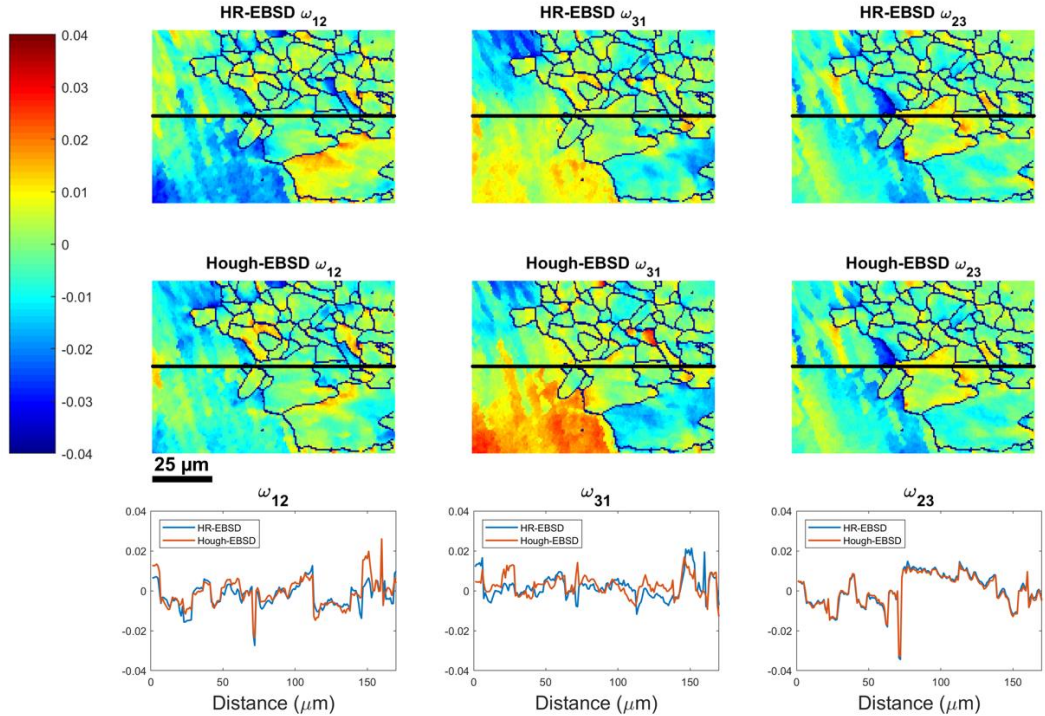

Fig. 8: Comparison of lattice rotation tensor components $\omega_{12}$, $\omega_{31}$ and $\omega_{23}$ (rad) obtained from HR-EBSD method and Hough-EBSD method for an as-built additively manufactured Inconel 625 sample.

The gradients in the lattice rotation can be further analyzed to map the geometrically necessary dislocation structures [36,37]. An energy minimization scheme is adopted to solve for FCC dislocation configurations [38], sufficient to accommodate the lattice curvature measured in HR-EBSD. Due to the difference in the angular resolution, the noise floor of the HR-EBSD approach is effectively an order of magnitude lower than the Hough-based approach. In Fig. 10(a-b), GND density maps obtained using both Hough-based and HR-EBSD based methods are plotted. The dislocation structure in the columnar grain is regularly aligned along the build direction. In addition, dislocation hot spots in the equiaxed region reside on similar locations to a deformed microstructure, close to geometric constraints like grain boundaries and triple junctions. Both methods capture the high density dislocation structures, whereas the low density dislocation structures are more clearly revealed in the HR-EBSD based method. It is reconfirmed by Fig.

10(c), in which the high dislocation density end of the two histograms match well, whereas the HR-EBSD extends to a much lower region of dislocation densities due to the enhanced sensitivity.

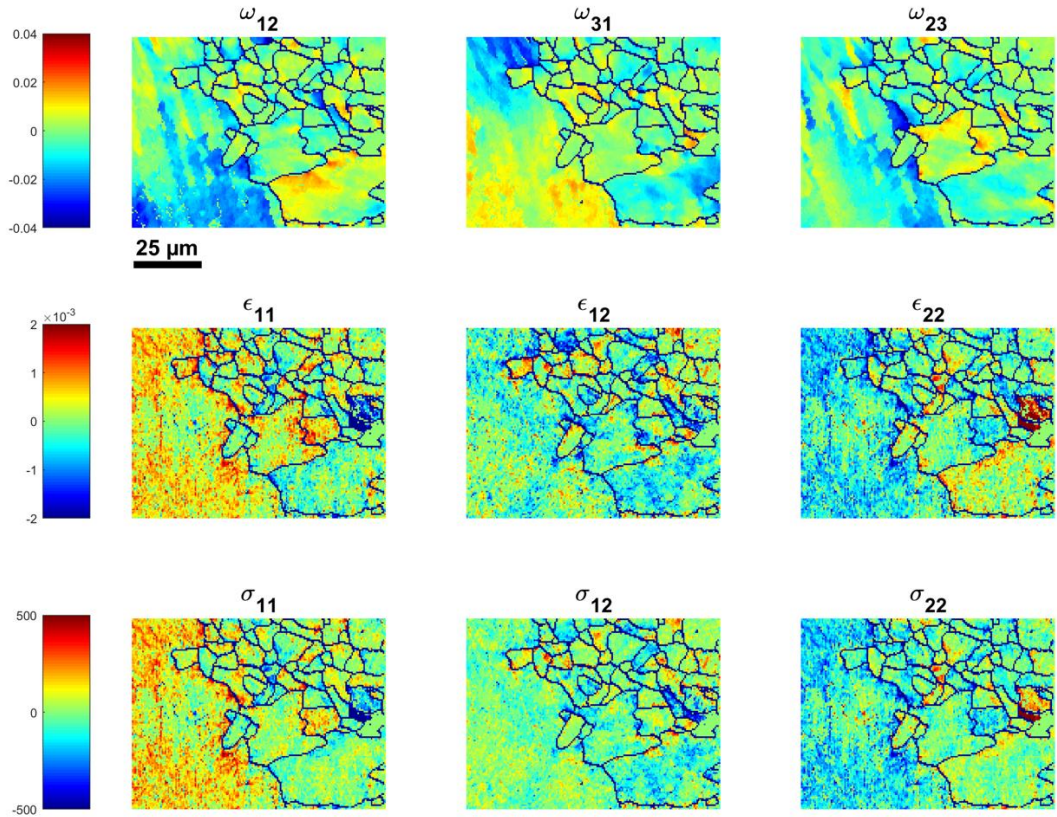

Fig. 9: (Top row) lattice rotation, (middle row) residual strain, and (bottom row) residual stress distribution for an as-built additively manufactured Inconel 625 sample.

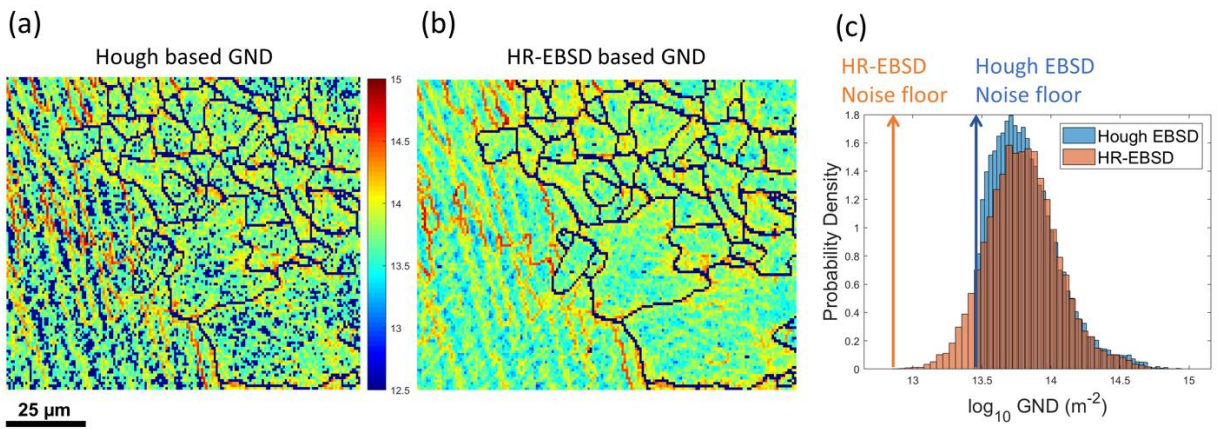

Fig. 10: (a) Hough based GND density map (log10 scale); (b) HR-EBSD based GND density map (log10 scale); (c) histograms of the dislocation density maps in (a) and (b).

## 4  Conclusion

It has been demonstrated in this study that the demons registration method is effective in remapping electron backscattered diffraction patterns for HR-EBSD applications. The proposed method is more robust to remap patterns of large misorientations compared to conventional cross-correlation remapping method at large misorientation angles. The fast and accurate first pass remapping based on demons registration at large misorientation angles eliminates the need for robust fitting or iterative remapping, which significantly improves computational speed without compromising the accuracy. The rotation, stress, and strain sensitivity of this approach are determined to be around $0.5\times10^{-4}$, 35MPa, and $2\times10^{-4}$, respectively. In addition, a case study on an additively manufactured Inconel 625 sample, using the developed method, reveals significant residual stress built-up in the columnar grain region and regularly aligned GND structures. Future development of this method includes optimization of the multi-resolution demons algorithm to eliminate the second-pass cross-correlation and incorporation of simulated patterns for measurement of absolute stress/strain.

Acknowledgement

C. Zhu would like to acknowledge Dissertation Year Fellowship provided by the home department. C. Zhu would also like to thank Dr. Ben Britton for fruitful discussions. K. Kaufmann acknowledges support by the Department of Defense (DoD) through the National Defense Science and Engineering Graduate Fellowship (NDSEG) Program. K. Kaufmann would also like to acknowledge the financial support of the ARCS Foundation.

# References

[1] A.J. Schwartz, M. Kumar, B.L. Adams, D.P. Field, Electron backscatter diffraction in materials science, New York: Kluwer Academic, 2000.

[2] I. Brough, P.S. Bate, F.J. Humphreys, Optimising the angular resolution of EBSD, Mater. Sci. Technol. 22 (2006) 1279–1286.

[3] K. Thomsen, N.H. Schmidt, A. Bewick, K. Larsen, J. Goulden, Improving the accuracy of orientation measurements using EBSD, Microsc. Microanal. 19 (2013) 724–725.

[4] C. Maurice, R. Fortunier, A 3D Hough transform for indexing EBSD and Kossel patterns, J. Microsc. 230 (2008) 520–529.

[5] A.J. Wilkinson, G. Meaden, D.J. Dingley, High-resolution elastic strain measurement from electron backscatter diffraction patterns: New levels of sensitivity, Ultramicroscopy. 106 (2006) 307–313.

[6] T.B. Britton, C. Maurice, R. Fortunier, J.H. Driver, A.P. Day, G. Meaden, D.J. Dingley, K. Mingard, A.J. Wilkinson, Factors affecting the accuracy of high resolution electron backscatter diffraction when using simulated patterns, Ultramicroscopy. 110 (2010) 1443–1453.

[7] J. Kacher, C. Landon, B.L. Adams, D. Fullwood, Bragg's Law diffraction simulations for electron backscatter diffraction analysis, Ultramicroscopy. 109 (2009) 1148–1156.

[8] T. Vermeij, M. De Graef, J. Hoefnagels, Demonstrating the potential of accurate absolute cross-grain stress and orientation correlation using electron backscatter diffraction, Scr. Mater. 162 (2019) 266–271.

[9] T. Tanaka, A.J. Wilkinson, High Angular Resolution Electron Backscatter Diffraction Studies of Tetragonality in Fe-C Martensitic Steels, Microsc. Microanal. 24 (2018) 962–963.

[10] T. Vermeij, J.P.M. Hoefnagels, A consistent full-field integrated DIC framework for HR-EBSD, Ultramicroscopy. 191 (2018) 44–50.

[11] T.B. Britton, A.J. Wilkinson, High resolution electron backscatter diffraction measurements of elastic strain variations in the presence of larger lattice rotations, Ultramicroscopy. 114 (2012) 82–95.

[12] C. Maurice, J.H. Driver, R. Fortunier, On solving the orientation gradient dependency of high angular resolution EBSD, Ultramicroscopy. 113 (2012) 171–181.

[13] F. Ram, S. Zaefferer, T. Jäpel, D. Raabe, Error analysis of the crystal orientations and disorientations obtained by the classical electron backscatter diffraction technique, J. Appl. Crystallogr. 48 (2015) 797–813.

[14] F.J. Humphreys, Grain and subgrain characterisation by electron backscatter diffraction, J. Mater. Sci. 36 (2001) 3833–3854.

[15] F.J. Humphreys, Quantitative metallography by electron backscattered diffraction, J. Microsc. 195 (1999) 170–185.

[16] S.I. Wright, M.M. Nowell, J. Basinger, Precision of EBSD based orientation measurements, Microsc. Microanal. 17 (2011) 406–407.

[17] T.B. Britton, A.J. Wilkinson, Measurement of residual elastic strain and lattice rotations with high resolution electron backscatter diffraction, Ultramicroscopy. 111 (2011) 1395–1404.

[18] T.J. Ruggles, G.F. Bomarito, R.L. Qiu, J.D. Hochhalter, New levels of high angular resolution EBSD performance via inverse compositional Gauss–Newton based digital image correlation, Ultramicroscopy. 195 (2018) 85–92.

[19] Q. Shi, S. Roux, F. Latourte, F. Hild, Estimation of Elastic Strain by Integrated Image Correlation on Electron Diffraction Patterns, Ultramicroscopy. 199 (2019) 16–33.

[20] B.K.P. Horn, B.G. Schunck, Determining optical flow, Artif. Intell. 17 (1981) 185–203.

[21] J.-M. Odobez, P. Bouthemy, Robust multiresolution estimation of parametric motion models, J. Vis. Commun. Image Represent. 6 (1995) 348–365.

[22] F. Hild, S. Roux, Comparison of Local and Global Approaches to Digital Image Correlation, Exp. Mech. 52 (2012) 1503–1519.

[23] P. Cheng, M.A. Sutton, H.W. Schreier, S.R. McNeill, Full-field speckle pattern image correlation with B-spline deformation function, Exp. Mech. 42 (2002) 344–352.

[24] S. Roux, F. Hild, Y. Berthaud, Correlation image velocimetry: a spectral approach, Appl. Opt. 41 (2002) 108–115.

[25] J.P. Thirion, Image matching as a diffusion process: An analogy with Maxwell's demons, Med. Image Anal. 2 (1998) 243–260.

[26] P.J. Kostelec, J.B. Weaver, D.M. Healy, Multiresolution elastic image registration, Med. Phys. 25 (1998) 1593–1604.

[27] A.J. Wilkinson, G. Meaden, D.J. Dingley, High resolution mapping of strains and rotations using electron backscatter diffraction, Mater. Sci. Technol. 22 (2006) 1271–1278.

[28] T.J. Hardin, T.J. Ruggles, D.P. Koch, S.R. Niezgoda, D.T. Fullwood, E.R. Homer, Analysis of traction-free assumption in high-resolution EBSD measurements, J. Microsc. 260 (2015) 73–85.

[29] E. Salvati, T. Sui, A.M. Korsunsky, Uncertainty quantification of residual stress evaluation

by the FIB-DIC ring-core method due to elastic anisotropy effects, Int. J. Solids Struct. 87 (2016) 61–69.

[30] P.G. Callahan, M. De Graef, Dynamical electron backscatter diffraction patterns. Part I: Pattern simulations, Microsc. Microanal. 19 (2013) 1255–1265.

[31] I. Campbell, D. Bourell, I. Gibson, Additive manufacturing: rapid prototyping comes of age, Rapid Prototyp. J. 18 (2012) 255–258.

[32] D. Herzog, V. Seyda, E. Wycisk, C. Emmelmann, Additive manufacturing of metals, Acta Mater. 117 (2016) 371–392.

[33] T. DebRoy, H.L. Wei, J.S. Zuback, T. Mukherjee, J.W. Elmer, J.O. Milewski, A.M. Beese, A. Wilson-Heid, A. De, W. Zhang, Additive manufacturing of metallic components–process, structure and properties, Prog. Mater. Sci. 92 (2018) 112–224.

[34] G.K. Lewis, E. Schlienger, Practical considerations and capabilities for laser assisted direct metal deposition, Mater. Des. 21 (2000) 417–423.

[35] P. Mercelis, J.-P. Kruth, Residual stresses in selective laser sintering and selective laser melting, Rapid Prototyp. J. 12 (2006) 254–265.

[36] C. Zhu, T. Harrington, V. Livescu, G.T. Gray, K.S. Vecchio, Determination of geometrically necessary dislocations in large shear strain localization in aluminum, Acta Mater. 118 (2016) 383–394.

[37] C. Zhu, T. Harrington, G.T. Gray, K.S. Vecchio, Dislocation-type evolution in quasi-statically compressed polycrystalline nickel, Acta Mater. 155 (2018) 104–116.

[38] E. Demir, D. Raabe, N. Zaafarani, S. Zaefferer, Investigation of the indentation size effect through the measurement of the geometrically necessary dislocations beneath small indents of different depths using EBSD tomography, Acta Mater. 57 (2009) 559–569.

**Supporting Material: Part 1**

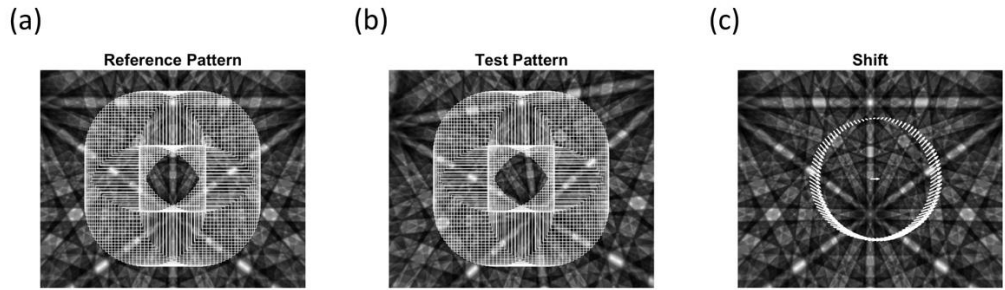

Fig. S1: (a) Reference pattern of fcc nickel with Euler angle [0,0,0] and 100 ROIs overlaid on to it; (b) test pattern of fcc nickel with Euler angle [10,0,0] and 100 ROIs overlaid on to it; (c) shifts $\vec{q}$ measured using cross-correlation between 100 ROIs in (a) and (b), and overlaid onto the reference pattern.

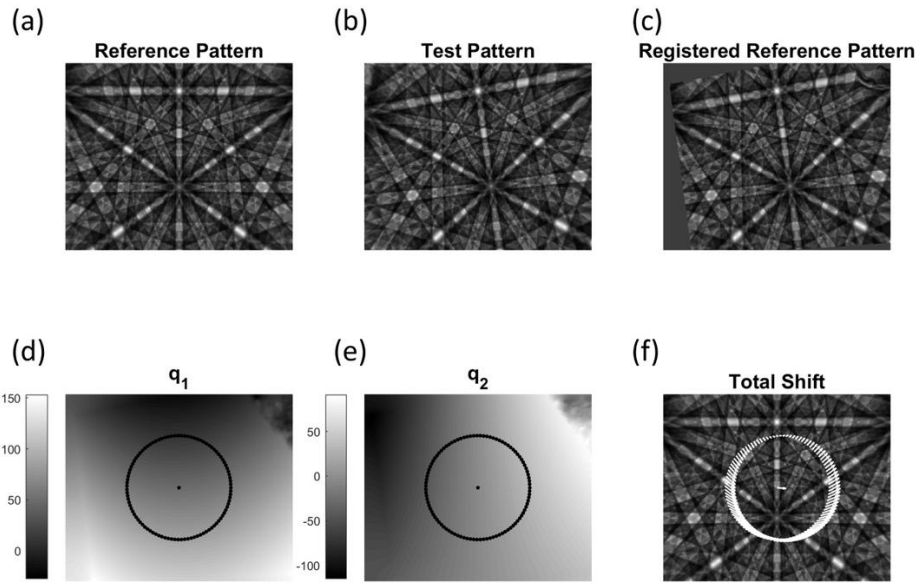

Fig. S2: (a) Reference pattern of fcc nickel with Euler angle [0,0,0] and 100 ROIs overlaid on to it; (b) test pattern of fcc nickel with Euler angle [10,0,0] and 100 ROIs overlaid on to it; (c) registered pattern with demons registration method; components of the displacements fields (d) $q_1$ and (e) $q_2$, which form the shift vectors (f) $\vec{q}=[\ q_1,\ q_2\ ]$ at the black dots (center of ROIs in Fig.A7.1).

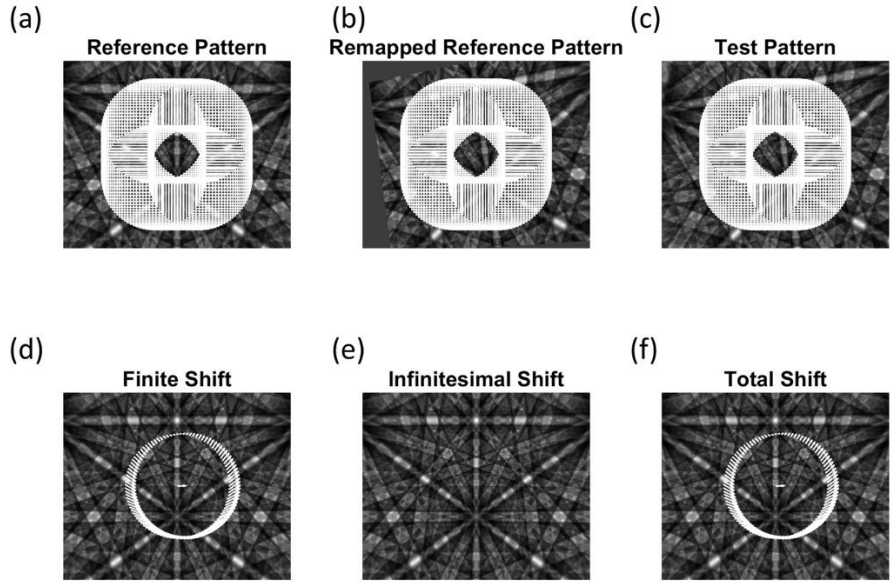

Fig. S3: (a) Reference pattern of fcc nickel with Euler angle [0,0,0] and 100 ROIs overlaid on to it; (b) test pattern of fcc nickel with Euler angle [10,0,0] and 100 ROIs overlaid on to it; (c) registered pattern with demons registration method; components of the displacements fields (d) $q_1$ and (e) $q_2$, which form the shift vectors (f) $\vec{q}$=[ $q_1$, $q_2$ ] at the black dots (center of ROIs in Fig.A7.1).

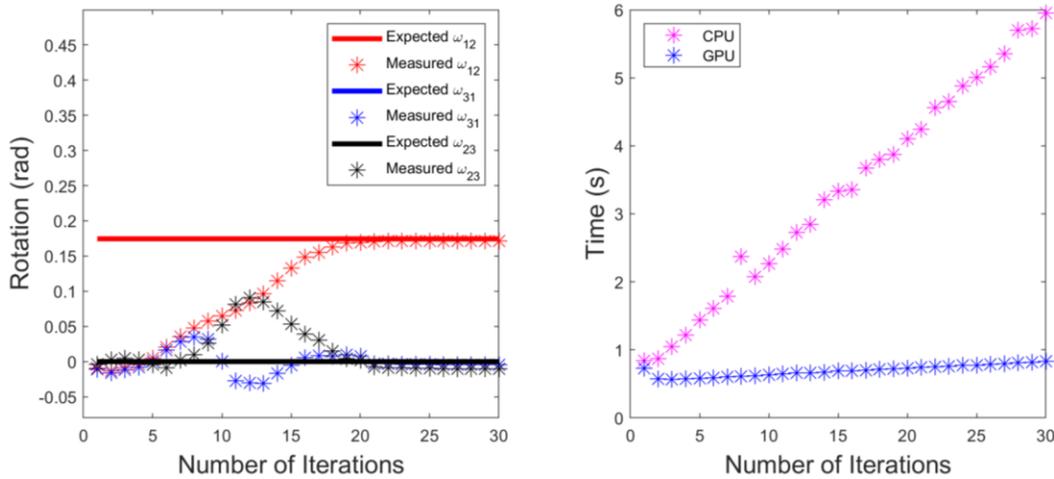

Fig. S4: Convergence of iterative multiresolution demons registration. (left) number of iterations (at each pyramid level) required register reference pattern with Euler angles [0,0,0] to test pattern with Euler angles [10,0,0]; (right) total time taken for the registration process using Intel i7-8750H CPU and NIVIDIA GeForce GTX 1050 Ti GPU.

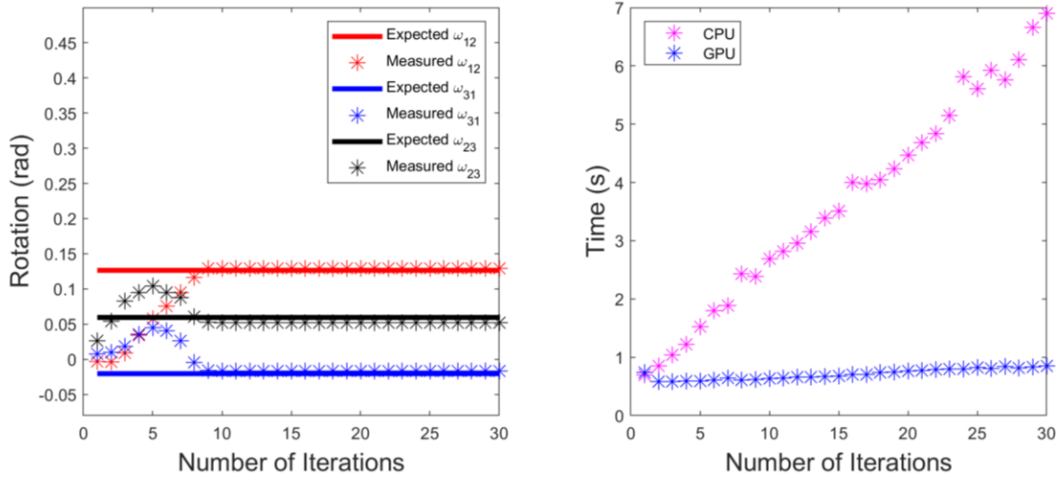

Fig. S5: Convergence of iterative multiresolution demons registration. (left) number of iterations (at each pyramid level) required register reference pattern with Euler angles [125°, 43°, 57°] to test pattern with Euler angles [130°,40°,60°]; (right) total time taken for the registration process using Intel i7-8750H CPU and NIVIDIA GeForce GTX 1050 Ti GPU.

**Supporting Material: Part 2**

Coordinate Transformation of Stiffness Matrix

In the Bunge notation (x-z-x), the orientation matrix **g** (passive) that brings the sample coordinates system coincident with crystal coordinate system is defined by multiplication of three rotation matrices:

$$\mathbf{g}(\varphi_2, \Phi, \varphi_1) = \mathbf{R}_x(\varphi_2)\mathbf{R}_z(\Phi)\mathbf{R}_x(\varphi_1) \tag{2.1}$$

$$\mathbf{R}_Z = \begin{pmatrix} \cos\theta & \sin\theta & 0 \\ -\sin\theta & \cos\theta & 0 \\ 0 & 0 & 1 \end{pmatrix} \tag{2.2}$$

$$\mathbf{R}_x = \begin{pmatrix} 1 & 0 & 0 \\ 0 & \cos\theta & \sin\theta \\ 0 & -\sin\theta & \cos\theta \end{pmatrix} \tag{2.3}$$

Since the orientation matrix is an orthogonal matrix, the reverse rotation matrix that brings the crystal coordinate system into alignment with sample coordinates system is the $\mathbf{g}^T$. Therefore, stress and strain tensor can be transformed from the local crystal coordinate system to the sample coordinate system:

$$\sigma_S = \mathbf{g}^T \sigma_C \mathbf{g} \tag{2.4}$$

$$\varepsilon_S = \mathbf{g}^T \varepsilon_C \mathbf{g} \tag{2.5}$$

Moreover, Hooke's law is defined more conveniently with the introduction of Voigt notation, which reduces the rank of stiffness tensor **C** (from fourth order to second order) by vectorizing the stress/strain tensors and considering the crystal symmetry information. By convention, the stiffness tensor C is typically expressed in the Cartesian crystal reference frame, which requires additional transformation operation to bring it into sample reference frame.

$$\vec{\sigma} = [\sigma_{11}, \sigma_{22}, \sigma_{33}, \sigma_{23}, \sigma_{13}, \sigma_{11}] \tag{2.6}$$

$$\vec{\varepsilon} = [\varepsilon_{11}, \varepsilon_{22}, \varepsilon_{33}, 2\varepsilon_{23}, 2\varepsilon_{13}, 2\varepsilon_{11}] \tag{2.7}$$

$$\vec{\sigma}_C = C\,\vec{\varepsilon}_C \tag{2.8}$$

The new transformation matrices, related to the components of **g**, to rotate the stress or strain vector from crystal frame to sample frame can be obtained by vectorizing Eqs. (2.4) - (2.5) and reorganizing the coefficients.

$$\vec{\sigma}_S = R_\sigma \vec{\sigma}_C \tag{2.9}$$

$$\vec{\varepsilon}_S = R_\varepsilon \vec{\varepsilon}_C \tag{2.10}$$

From Hooke's law in Eq. (2.8), the sample vector stress $\vec{\sigma}_S$ can be related to the sample vector strain $\vec{\varepsilon}_S$ through the stiffness matrix in the sample frame $R_\sigma C R_\varepsilon^{-1}$.

$$\vec{\sigma}_S = R_\sigma \vec{\sigma}_C = R_\sigma C \vec{\varepsilon}_C = R_\sigma C R_\varepsilon^{-1} \vec{\varepsilon}_S \tag{2.11}$$

The surface traction free boundary condition is then defined by setting the corresponding surface normal stress $\vec{\sigma}_S(3)$ and shear stresses $\vec{\sigma}_S(4), \vec{\sigma}_S(5)$ to be zeroes.